\documentclass[runningheads]{CMSIM}

\usepackage{graphicx} 

\usepackage{amsmath,amssymb}

\renewcommand{\thefootnote}

\title*{Delay as an energy regulator of the generation of deterministic chaos in hydrodynamic systems with limited excitation}

\titlerunning{\it Running Paper Title}

\author{
Aleksandr Shvets$^1$
\and
  Ilmi Seit-Dzhelil$^2$
}

\authorrunning{\it Shvets and  Seit-Dzhelil}

\institute{
$^1$
National Academy of Sciences of Ukraine, Institute of Mathematics, Kyiv, Ukraine\\
(E-mail: {\tt oshvets@imath.kiev.ua})\\
$^2$
National Technical University of Ukraine “Igor Sikorsky Kyiv Polytechnic Institute”, Kyiv, Ukraine\\
(E-mail: {\tt ilmiseitdzelil17@gmail.com})
}

\begin{document}
\thispagestyle{empty}
\maketitle
\setlength{\leftskip}{0pt}
\setlength{\headsep}{16pt}
\begin{abstract}
The Miles-Krasnopolskaya system is considered, which is used to study the nonlinear interaction of a tank with a liquid and the source of excitation of its oscillations. Additionally, delay time of impulse from the source of excitation of oscillations on the dynamics of the aggregate system "tank with liquid - source of excitation" is taken into account. A technique for studying the attractors of such systems is proposed.

It is shown that delay plays a key role in the emergence (disappearance) of deterministic chaos in the Miles-Krasnopolskaya system. Quantitative changes in the value of the delay can lead to qualitative changes in the types of attractors of the system. So, regular attractors can turn into chaotic ones and vice versa. Also, a change in the delay value can lead to the implementation of new scenarios, both transitions from regular attractors to chaotic ones and transitions from a chaotic attractor of one type to a chaotic attractor of another type.
\end{abstract}

\section{Introduction}

The study of vibrations of the free surface of a liquid in rigid tanks is considered in a large number of works, a detailed bibliography of which is given in monographs Ibrahim \cite{Ibr}, Lukovsky \cite{Luk}. In addition to great research interest, these problems have wide practical application in many areas of modern technology, since many modern machines, mechanisms and vehicles incorporate tanks of various shapes that are partially filled with liquid as integral structural components.
In the vast majority of works, the fluctuations of liquid in tanks are considered in the so-called ``ideal'' formulation. With this formulation of the problem, it is assumed that the source of excitation of liquid vibrations has enormous power. As a result, the reverse effect of the oscillatory system, in this case a tank partially filled with liquid, on the source of excitation of oscillations can be neglected. The urgent need for global energy saving requires maximum minimization of the power of the embedded sources of excitation of oscillations. This leads to
the fact that the power of used excitation sources becomes comparable to the power consumed by the oscillatory system. This is exactly the situation that most often occurs in real machines and mechanisms. In such cases, the use of ``ideal'' mathematical models can lead to gross errors in the description of dynamic systems ``excitation source --  oscillatory subsystem''. In particular, information about the existence of deterministic chaos in such systems may be lost (Frolov and Krasnopolskaya \cite{kr1}, Krasnopolskaya and Shvets \cite{kr2}, \cite{kr-sh}, Shvets \cite{s25}).

Another important factor influencing the dynamics may be various delays of impacts. Note that the delay is always present in real systems due to the limited speed of signal transmission: compression and tension waves, current strength and electrical voltage, as well as many other factors. In some cases, taking into account the delay of impacts does not lead to significant changes in the dynamic characteristics of the systems under study. In others cases, the presence of a delay leads not only to significant quantitative changes in parameters of steady-state modes movement (interaction), but completely qualitatively changes the type of movements being studied (Shvets and Makasyeyev \cite{sh-m}, Shvets and Donetskyi \cite{sh-d}).

\section{Main mathematical model}

Let us consider dynamic system consisting of an electric motor of limited power and a cylindrical tank partially filled with liquid. Assume that an electric motor using a crank mechanism, excites horizontal vibrations of the platform of a tank with liquid. The scheme of such a system is shown in Fig. 1. Such a dynamic system is non-ideal by Sommerfeld-Kononenko (\cite{som}, \cite{kon}).
\begin{figure}
  \centerline{
    \includegraphics[width=0.75\textwidth]{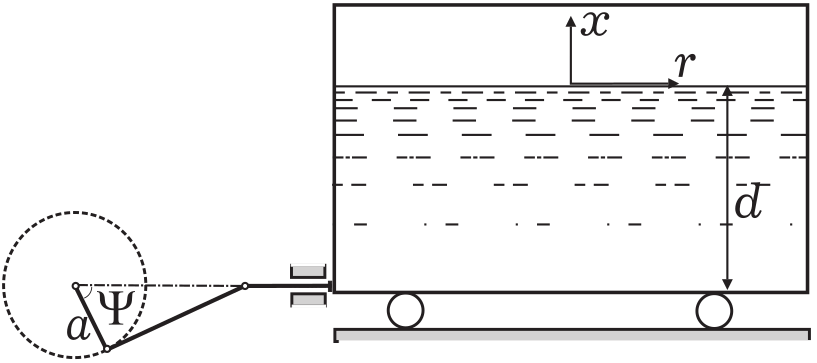}}
  \caption{Scheme of the system ``tank with liquid--electric motor" }
  \label{fig:0}
\end{figure}
The rotation of the electric motor shaft can be described by knowing changing of angle $\Psi$. When the crank $a$ is turned to an angle $\Psi$, the tank receives a movement in the horizontal plane $u(t)=a\cos\Psi(t)$. Assume that the liquid in the tank is inviscid, incompressible and has density $\rho$. We denote the radius of the cylindrical tank by $R$, and its cross section by $S$.

Suppose that liquid fills the tank to  depth $x=-d$. We only consider the direction of movements $u(t)$ of the tank platform which coincide with the direction of the polar axis $Or$. To describe the oscillations of the free surface of the liquid, we introduce the cylindrical coordinate system $Oxr\Psi$ with the origin on the axis of symmetry of the tank, on the undisturbed surface of the liquid. The equation of the relief of the free surface of the liquid can be written in the form $x=\zeta(r,\Psi,t)$

We will describe the movement of the liquid in the tank using the potential of the liquid velocity $\varphi(x,r,\Psi,t)$.

Let us represent the functions $\zeta(r,\Psi,t)$ and $\varphi(x,r,\Psi,t)$ in the form of series according to the eigenmodes of oscillations:
\begin{equation}
\begin{array}{c}
\displaystyle \zeta(r,\theta,t)=\sum_{i\,j}[
q_{i\,j}^c(t)\chi_{ij}(r)\cos{i\theta}+q_{i\,j}^s(t)\chi_{ij}(r)\sin{i\theta}
];\\[8mm]
\displaystyle \varphi(x,r,\theta,t)=\sum_{i\,j}\left[
\varphi_{i\,j}^c (t) X_{i\,j}(x,r) \cos{i \theta}+ \varphi_{i\,j}^s
(t) X_{i\,j}(x,r) \sin{i \theta} \right] \label{eq:book5.4},
\end{array}
\end{equation}
where $q_{i\;j}^c$, $q _{i\;j}^s$ and $\varphi _{i\;j}^c$, $\varphi_{i\;j}^s$~-- unknown amplitudes of normal modes; $\chi_{ij}(r)\cos{i\theta}$, $\chi_{ij}(r)\sin{i\theta}, X_{i\,j}$, $
\cos{i \theta}$ and $X_{i\,j} \sin{i \theta}$~-- eigenmodes in a linear approximation of the problem of oscillations of an ideal fluid in a cylindrical shell.

Based on the results of the work of Miles (\cite{Miles1976}--\cite{Miles1984c}) it was shown in the works of Krasnopolskaya and Shvets (\cite{kr-sh11}, \cite{kr-sh12}) that the process of nonlinear interaction between oscillations of the free surface of a liquid in a cylindrical tank by the main resonant modes and the rotation of the shaft of an electric motor of limited power can be described by a system of five differential equations:
\begin{equation}\label{eq1}
    \begin{aligned}
        &\frac{dp_1(\tau)}{d\tau} = \alpha p_1(\tau) - [\beta(\tau - \delta) + \frac{A}{2}(p_{1}^2(\tau) + q_{1}^2(\tau) + p_{2}^2(\tau) + q_{2}^2(\tau))]q_1(\tau)+\\ &+ B(p_{1}(\tau)q_{2}(\tau) - p_{2}(\tau)q_{1}(\tau))p_2(\tau); \\
        &\frac{dq_1(\tau)}{d\tau} = \alpha q_1(\tau) + [\beta(\tau - \delta) + \frac{A}{2}(p_{1}^2(\tau) + q_{1}^2(\tau) + p_{2}^2(\tau) + q_{2}^2(\tau))]p_1(\tau)+\\ &+ B(p_{1}(\tau)q_{2}(\tau) - p_{2}(\tau)q_{1}(\tau))q_2(\tau) + 1;\\
        &\frac{d\beta(\tau)}{d\tau} = N_3 - \mu_1 q_1(\tau) + N_1\beta(\tau); \\
        &\frac{dp_2(\tau)}{d\tau} = \alpha p_2(\tau) - [\beta(\tau - \delta) + \frac{A}{2}(p_{1}^2(\tau) + q_{1}^2(\tau) + p_{2}^2(\tau) + q_{2}^2(\tau))]q_2(\tau)-\\ &- B(p_{1}(\tau)q_{2}(\tau) - p_{2}(\tau)q_{1}(\tau))p_1(\tau);\\
        &\frac{dq_2(\tau)}{d\tau} = \alpha q_2(\tau) + [\beta(\tau - \delta) + \frac{A}{2}(p_{1}^2(\tau) + q_{1}^2(\tau) + p_{2}^2(\tau) + q_{2}^2(\tau))]p_2(\tau)-\\ &- B(p_{1}(\tau)q_{2}(\tau) - p_{2}(\tau)q_{1}(\tau))q_1(\tau).
    \end{aligned}
\end{equation}

Here, the phase variables $p_1, q_1$ and $p_2, q_2$ are the amplitudes of  oscillations of free surface of the liquid by the first and second main dominant modes, respectively; the phase variable $\beta$ is proportional to the rotation speed of the electric motor shaft; $ \tau$ is dimensionless time; $\alpha$ is a coefficient of viscous damping forces; $\mu_1$ is a coefficient of proportionality of vibration moment; $N_1$ is a angle of inclination of the static characteristic of the electric motor; parameters $A$ and $B$ are constants that depend on  radius of the tank $R$ and the height of the liquid $-d$ poured into it: $N_3$ is a multiparameter that depends on the radius of the tank, the length of the crank and eigen frequency of the mainl tone of free surface oscillations (\cite{kon}, \cite{Miles1976}--\cite{kr-sh12}).

Particular attention should be given to the constant $\delta$, a non-negative value that denotes the delay in the impact of the electric motor impulse on the tank partially filled with liquid. The value $\delta$ represents the delay of the reaction when the signal from electric motor passes through the connecting rod, link joints and other connecting elements.  Especially the delay must be taken into account when the source of excitement is remote from the tank.

The system of equations (\ref{eq1}) is a nonlinear system with a delayed argument. In general, the system of equations (\ref{eq1}) belongs to one of the types of Miles-Krasnopolskaya systems.

Constructing solutions to systems of equations with delay is in many cases a very difficult task. Therefore, we will attempt to simplify the system of equations $(\ref{eq1})$. Let us assume that the delay $\delta$ is small and expand the function $\beta(\tau-\delta)$ in the Maclaurin series by powers of small delay:
\begin{equation}\label{eq2}
    \beta(\tau - \delta) = \beta(\tau) - \delta \cdot \frac{d\beta(\tau)}{d\tau} +\delta^2  ...
\end{equation}

Substituting the obtained expressions for $\beta(\tau - \delta)$ into the system of equations (\ref{eq1}) obtained, up to terms of second order of smallness
\begin{equation}\label{eq3}
    \begin{aligned}
        &\frac{dp_1(\tau)}{d\tau} = \alpha p_1(\tau) - [\beta(\tau) - \delta\{N_3 - \mu_1 q_1(\tau) + N_1\beta(\tau)\} + \frac{A}{2}(p_{1}^2(\tau) +\\&+ q_{1}^2(\tau) + p_{2}^2(\tau) + q_{2}^2(\tau))]q_1(\tau)+ B(p_{1}(\tau)q_{2}(\tau) - p_{2}(\tau)q_{1}(\tau))p_2(\tau); \\
        &\frac{dq_1(\tau)}{d\tau} = \alpha q_1(\tau) + [\beta(\tau) - \delta\{N_3 - \mu_1 q_1(\tau) + N_1\beta(\tau)\} + \frac{A}{2}(p_{1}^2(\tau) +\\&+ q_{1}^2(\tau) + p_{2}^2(\tau) + q_{2}^2(\tau))]p_1(\tau) + B(p_{1}(\tau)q_{2}(\tau)- p_{2}(\tau)q_{1}(\tau))q_2(\tau)+ 1;\\
        &\frac{d\beta(\tau)}{d\tau} = N_3 - \mu_1 q_1(\tau) + N_1\beta(\tau); \\
        &\frac{dp_2(\tau)}{d\tau} = \alpha p_2(\tau) - [\beta(\tau) - \delta\{N_3 - \mu_1 q_1(\tau) + N_1\beta(\tau)\} + \frac{A}{2}(p_{1}^2(\tau) +\\&+ q_{1}^2(\tau) + p_{2}^2(\tau) + q_{2}^2(\tau))]q_2(\tau) - B(p_{1}(\tau)q_{2}(\tau) - p_{2}(\tau)q_{1}(\tau))p_1(\tau);\\
        &\frac{dq_2(\tau)}{d\tau} = \alpha q_2(\tau) + [\beta(\tau) - \delta\{N_3 - \mu_1 q_1(\tau) + N_1\beta(\tau)\} + \frac{A}{2}(p_{1}^2(\tau) +\\&+ q_{1}^2(\tau) + p_{2}^2(\tau) + q_{2}^2(\tau))]p_2(\tau) - B(p_{1}(\tau)q_{2}(\tau) - p_{2}(\tau)q_{1}(\tau))q_1(\tau).
    \end{aligned}
\end{equation}
The system of equations (\ref{eq3}) is a system of ordinary differential equations, which includes the delay as an additional parameter. The transformed system of equations (\ref{eq3}) retains its nonlinearity, necessitating the use of various numerical methods and algorithms for constructing solutions in the general case. The application of these methods to non-ideal dynamic systems is detailed in the works of  Krasnopolskaya and Shvets \cite{kr-sh12}  and Shvets and Krasnopolskaya \cite{sh-kr}.

\section{Numerical results}

The presence of a delay in the system leads to an adjustment of the energy interaction between the source of excitation of oscillations (electric motor) and the oscillatory system (tank partially filled with liquid). Changing the delay value can lead to both the appearance of chaotic attractors and the transformation of chaotic attractors into regular ones. Let us demonstrate such a process for concrete parameter values.

Let us assume that the system parameters (\ref{eq3}) are equal:
\begin{equation}
\alpha = -0.1,   \mu_1 = 0.5,  N_1 = -0.271, N_3=-0.1,  A = 1.12,  B = -1.531.
\label{set1}
\end{equation}
We will henceforth call this set of parameters by the first set of parameters.

\begin{figure}[hb]
  \centerline{
    \includegraphics[width=0.55\textwidth]{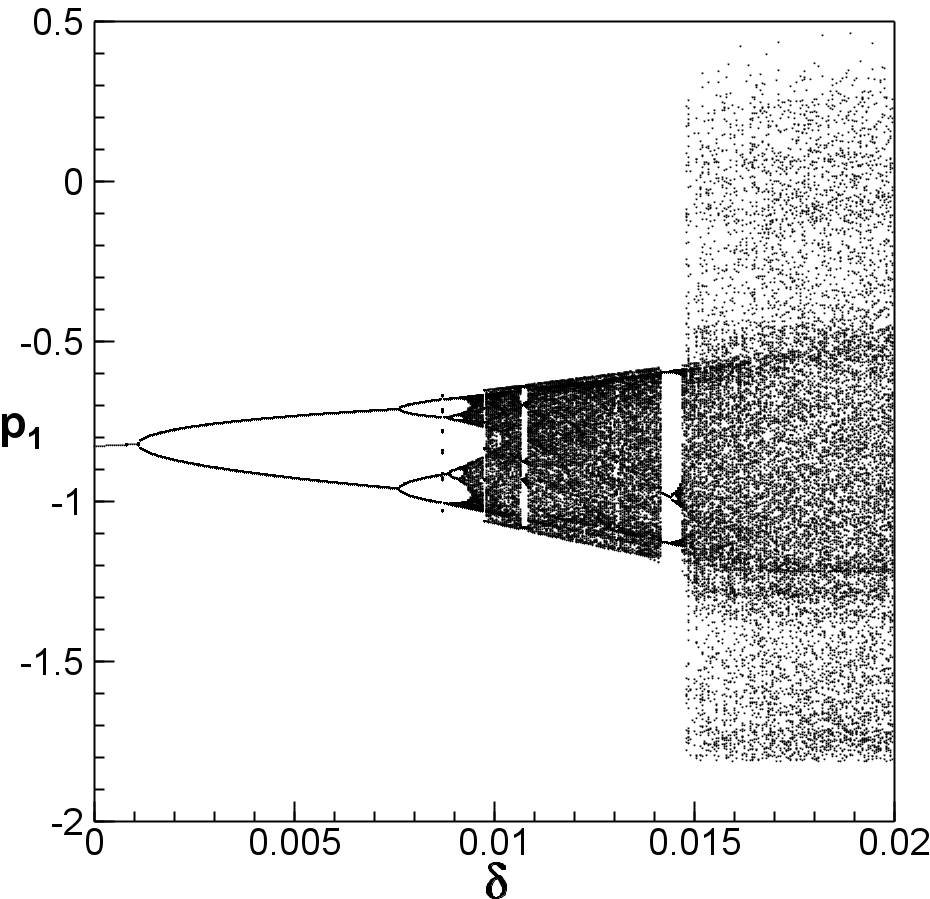}}
  \caption{Phase-parametric characteristic for first set of parameters}
  \label{fig:1}
\end{figure}

In Fig. \ref{fig:1} the phase-parametric characteristic of the system (\ref{eq3}) was built for the first set of parameters. To construct this phase-parametric characteristic, H\`{e}non method with a secant plane $\beta=-1.35$ was used. The constructed bifurcation tree makes it possible to analyze changes in the types of attractors of the system when the delay values change. Thus, in the absence of delay $\delta=0$, the attractor of the system is the limit cycle. Further, with increasing delay values, a cascade of period doubling bifurcations begins, which leads to the emergence of a chaotic attractor at $\delta\approx 0.009$ (Feigenbaum \cite{fei}). Let us recall that individual branches of the bifurcation tree correspond to limit cycles, and densely black regions on this tree correspond to chaotic attractors. Also in this figure, windows of periodicity in regions of chaos are clearly visible.

\begin{figure}[ht]
\centering
\begin{minipage}{0.5\textwidth}
\centering
\includegraphics[width=\textwidth] {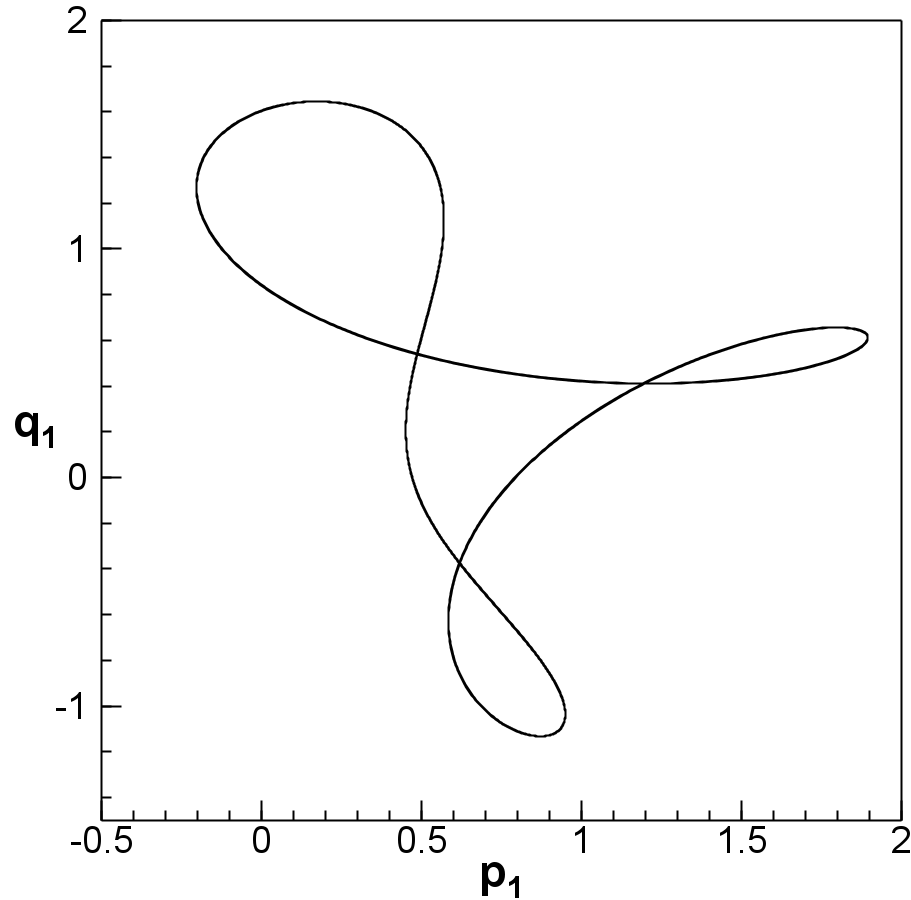}
(a)
\end{minipage}\hfill
\begin{minipage}{0.5\textwidth}
\centering
\includegraphics[width=\textwidth] {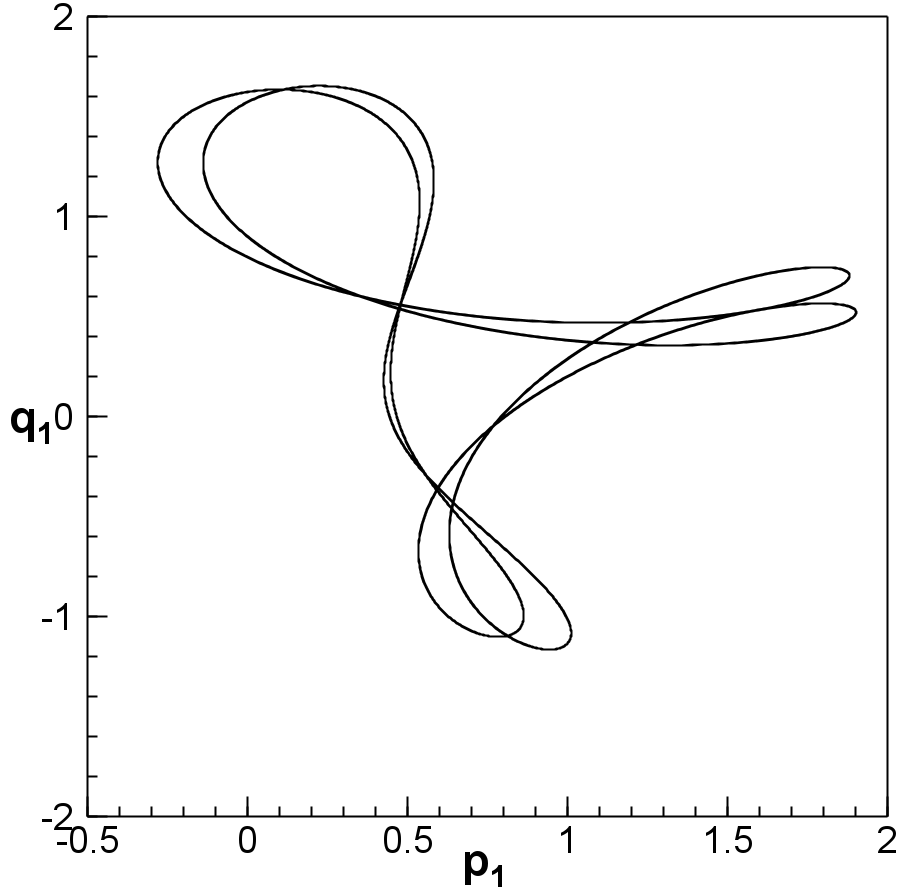}
(b)
\end{minipage}
\begin{minipage}{0.5\textwidth}
\centering
\includegraphics[width=\textwidth] {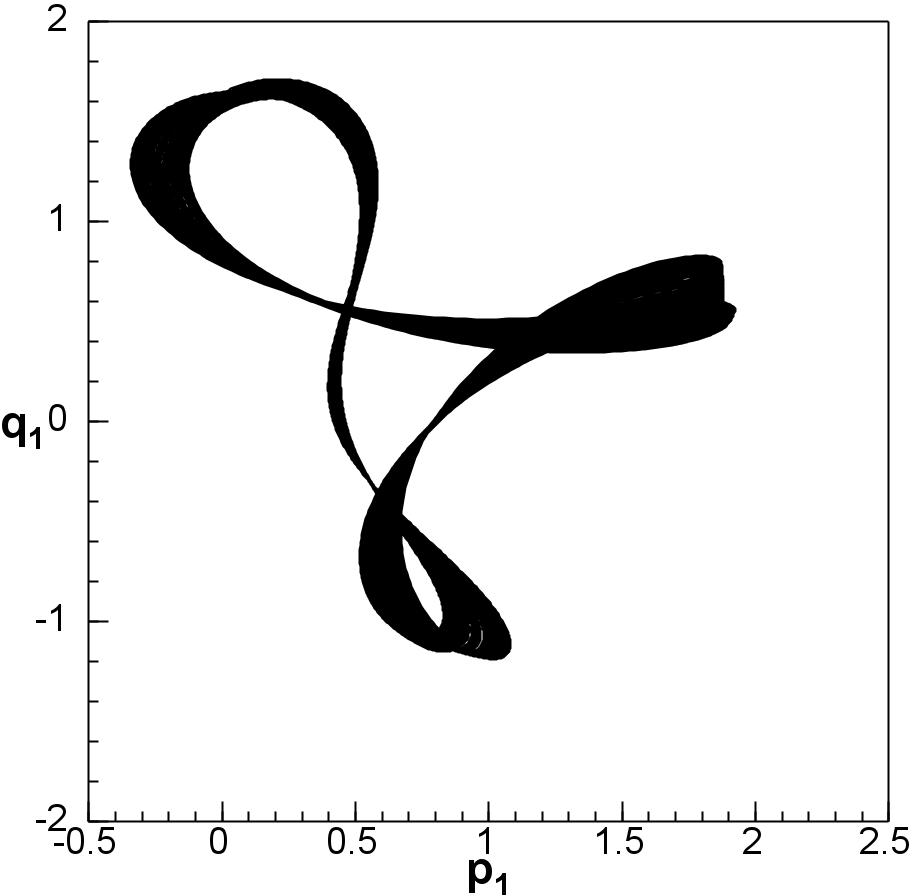}
(c)
\end{minipage}\hfill
\begin{minipage}{0.5\textwidth}
\centering
\includegraphics[width=\textwidth] {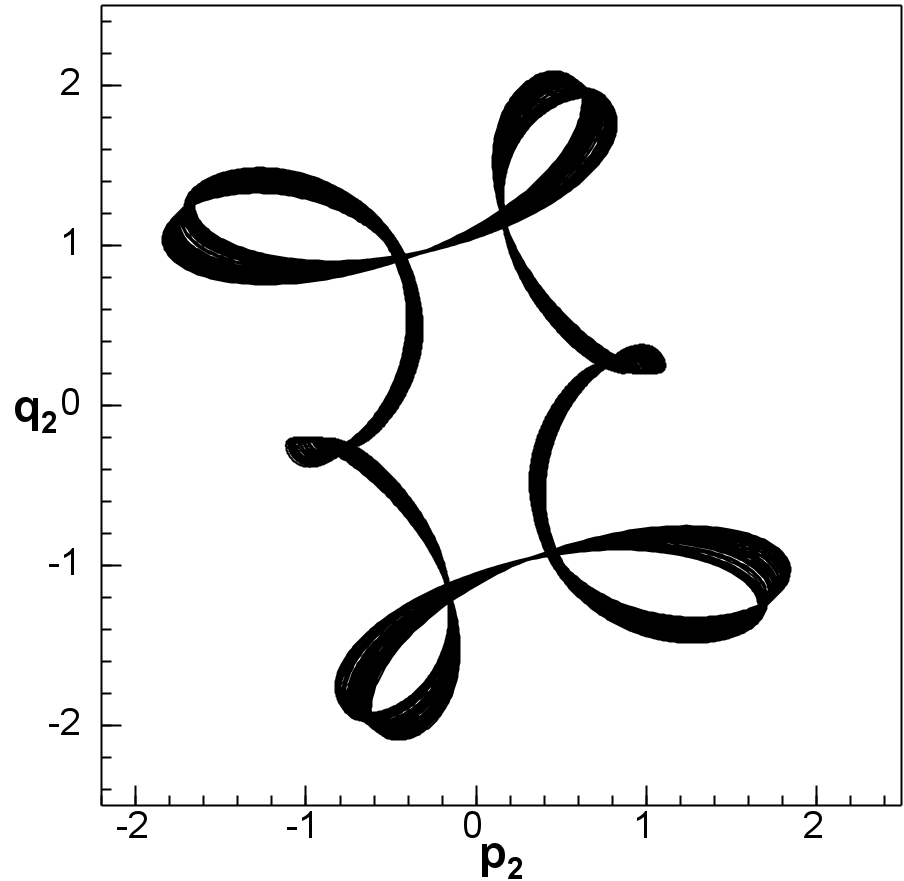}
(d)
\end{minipage}
\caption{Limit cycles at $\delta =0$ (a) and at $\delta =0.0075$ (b); Chaotic atttractors at $\delta =0.0099$  (c), (d)}\label{fig:2}
\end{figure}

In Fig. \ref{fig:2} projections of phase portraits were constructed for some values of delay. At $\delta=0$ (Fig. \ref{fig:2} (a)) the attractor of the system is the limit cycle of a simple one-turn structure. The limit cycle constructed at $\delta=0.0075$ (Fig. \ref{fig:2} (b)) corresponds to the first bifurcation of the period doubling. Finally, in (Fig. \ref{fig:2} (c), (d)) two projections of the chaotic attractor that exists in the system (\ref{eq3}) at $\delta=0.0099$ are shown.

Note that for all the considered attractors, the spectra of Lyapunov characteristic exponents were calculated. It has been determined, that for all limit cycles the maximum Lyapunov characteristic exponent is equal to zero, while chaotic attractors have a maximum Lyapunov characteristic exponent greater than zero.
Such values of characteristic  exponents are additional proof of the periodicity or chaotic nature of the corresponding attractor.

\begin{figure}[hb]
  \centerline{
    \includegraphics[width=0.55\textwidth]{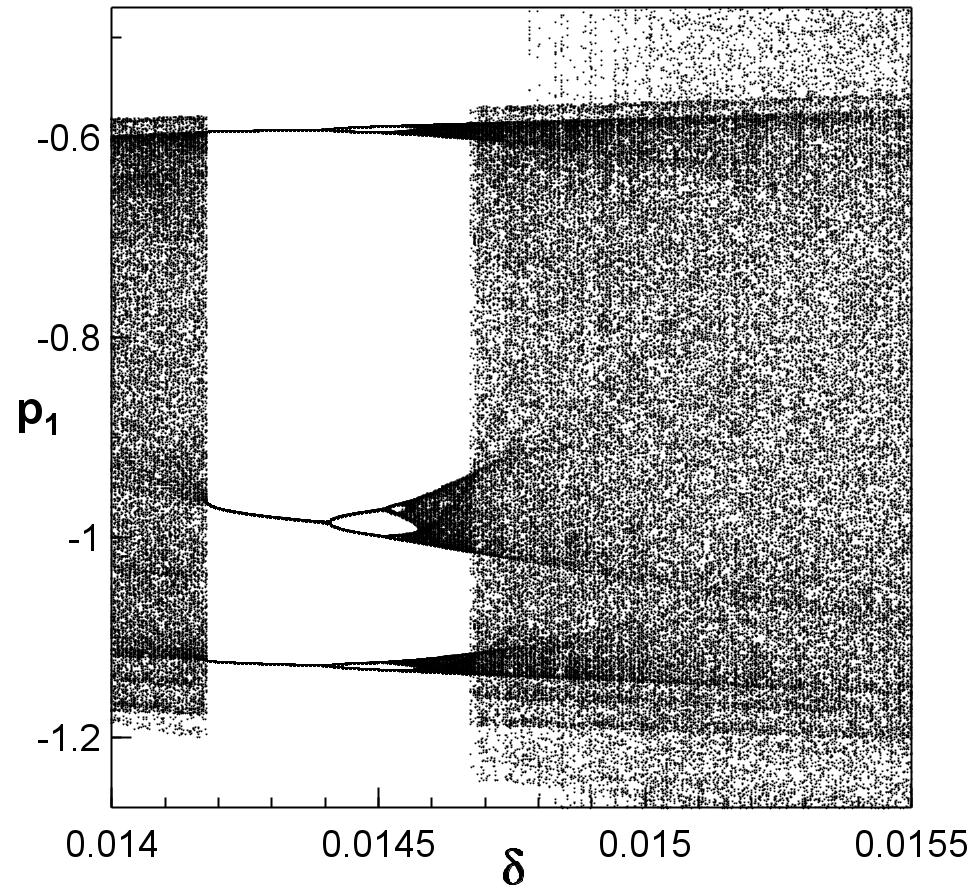}}
  \caption{Fragment of phase-parametric characteristic}
  \label{fig:3}
\end{figure}

In Fig. \ref{fig:3} shown an enlarged fragment of the phase-parametric characteristic, which corresponds to one of the periodicity windows. At the left border of this window, the attractor of the system is the limit cycle. As the delay value increases the cascade of period doubling bifurcations occurs. This endless cascade ends with the emergence of a chaotic attractor at $\delta\approx 0.0146$. With a further increase in the delay, the existing chaotic attractor disappears and a chaotic attractor of a new type appears in the system. The emergence of a new chaotic attractor occurs by one rigid bifurcation according to the scenario of generalized intermittency (Krasnopolskaya and Shvets \cite{kr-sh12}, Shvets \cite{s25}). A geometric indication of the presence of such a scenario is a significant increase in the area of the densely black (chaotic) region on the bifurcation tree.

\begin{figure}[ht]
\centering
\begin{minipage}{0.5\textwidth}
\centering
\includegraphics[width=\textwidth] {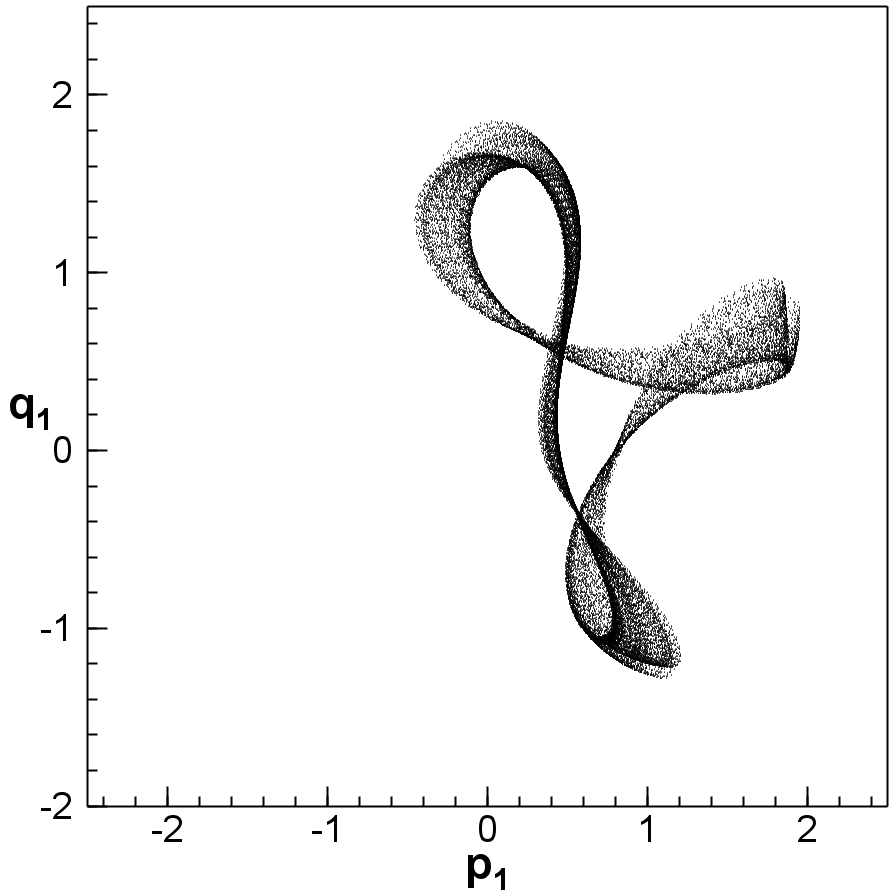}
(a)
\end{minipage}\hfill
\begin{minipage}{0.5\textwidth}
\centering
\includegraphics[width=\textwidth] {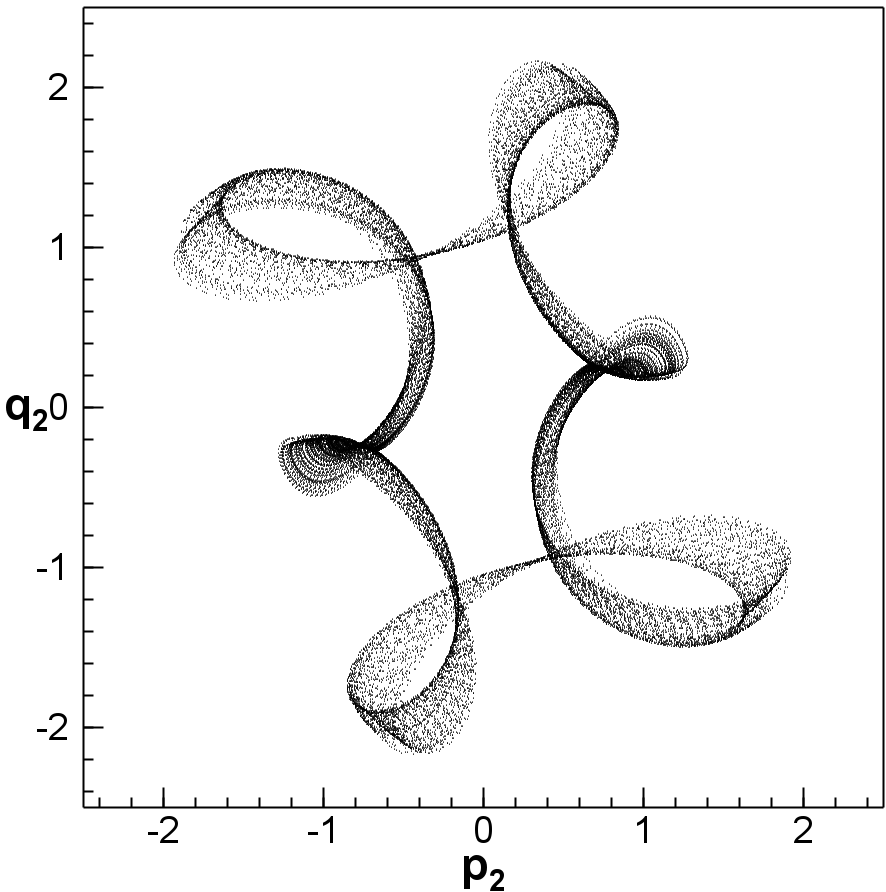}
(b)
\end{minipage}
\begin{minipage}{0.5\textwidth}
\centering
\includegraphics[width=\textwidth] {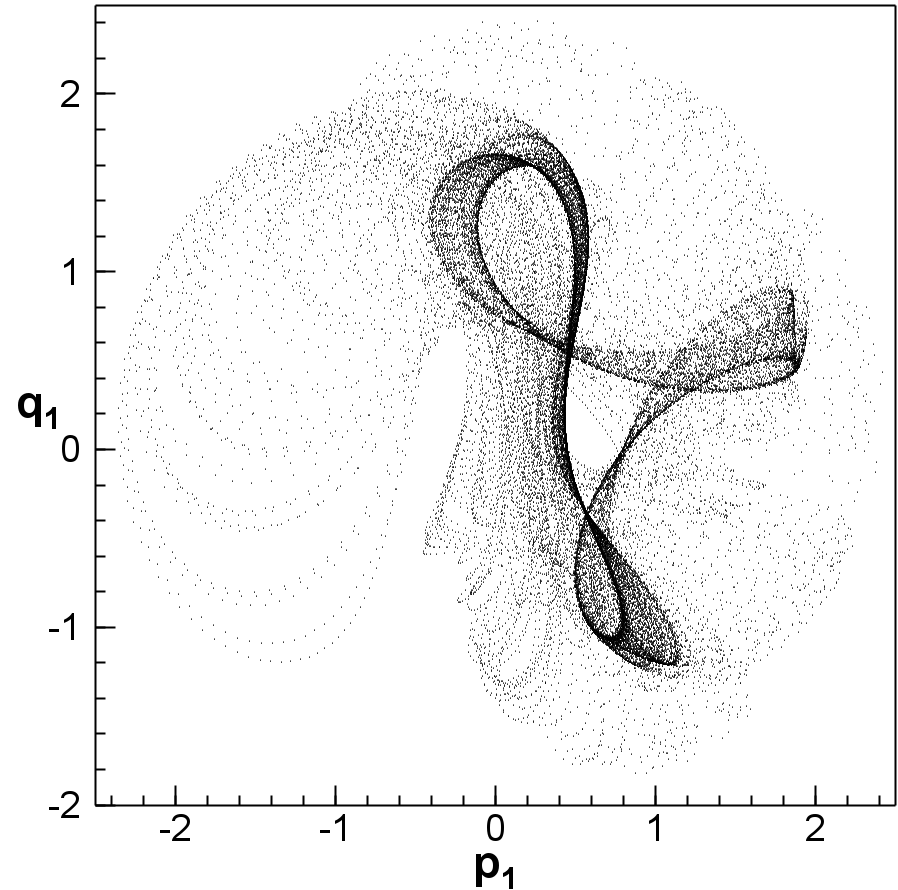}
(c)
\end{minipage}\hfill
\begin{minipage}{0.5\textwidth}
\centering
\includegraphics[width=\textwidth] {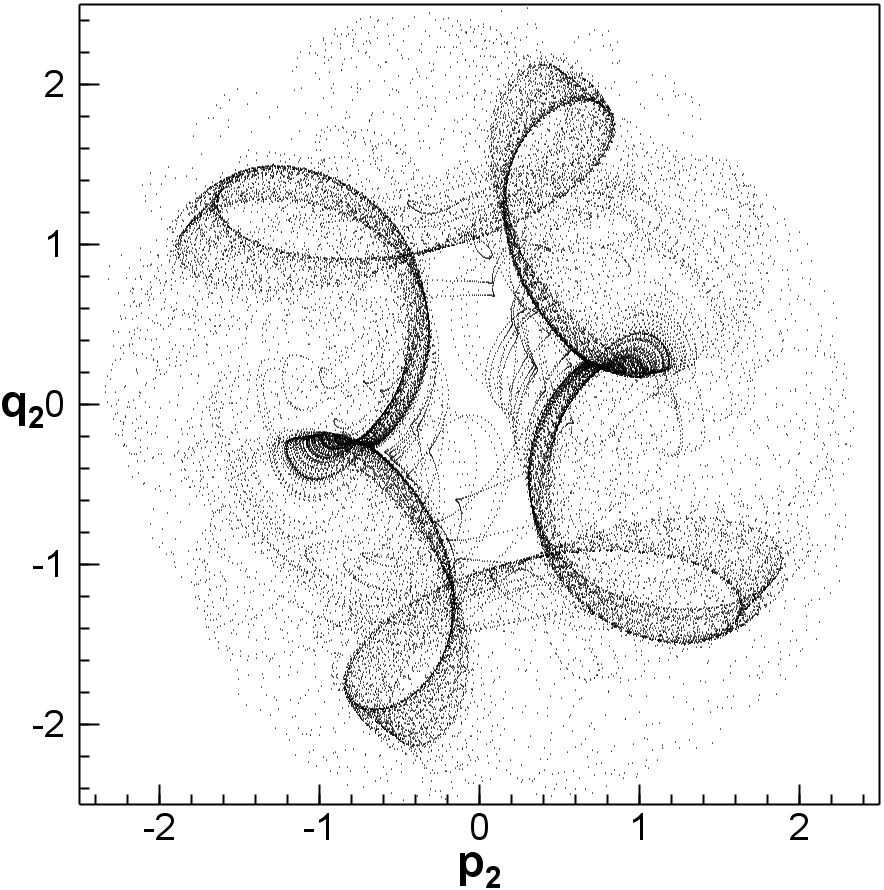}
(d)
\end{minipage}
\caption{Chaotic attractor at $\delta=0.01479$ $(a) ,(b)$; chaotic attractor at $\delta=0.0148$ $(c) ,(d)$;} \label{fig:4}
\end{figure}

In Fig. \ref{fig:4} distributions of the natural invariant measure of projections of phase portraits were constructed for two types of chaotic attractors existing in the system before and after the bifurcation point $\delta\approx 0.014795 $. A chaotic attractor of the first type exists in a relatively small interval of changes in delay values $0.0146<\delta< 0.014795.$ In addition, this attractor has a relatively small localization region in phase space. On the other hand, chaotic attractor of the second type exists for a much larger interval of changes in delay values $0.014975<\delta<0.016$ and has a significantly larger localization area in phase space. So, as the delay value increases after passing the bifurcation point, the chaotic attractor of the first type disappears and a chaotic attractor of the second type appears in the system. The movement of trajectories belonging to a chaotic attractor of the second type consists of two phases: rough laminar and turbulent. In the rough laminar phase (significantly more densely plotted points in Fig. \ref{fig:4} (c), (d)) the trajectory makes chaotic wanderings in the region of localization in the phase space of the chaotic attractors of the first type, which disappears after such a bifurcation. Note that the distribution regions of the invariant measure in the rough laminar phase practically coincide with the distribution regions of the invariant measure of the disappeared chaotic attractor of the first type. At an unpredictable moment in time, the trajectory abruptly departs from this region to more distant parts of the phase space. Such a phase of intermittency is called turbulent. The turbulent phase corresponds to much less frequently plotted points in Fig. \ref{fig:4} (c), (d). This process of changing the phases of trajectories is repeated an infinite number of times. Moreover, the transition time from rough laminar phase into turbulent phase
and back to rough laminar is unpredictable.

It should be noted that another evidence of the implementation of the generalized intermittency scenario is a significant increase of the maximum Lyapunov characteristic exponent of a chaotic attractor of the second type compared to the similar Lyapunov exponent of a chaotic attractor of the first type. Moreover, such an increase is observed immediately after passing the bifurcation point.

Let us consider another example of the influence of delay on the type of attractors of the "electric motor - tank partially filled with liquid" system. Assume that the parameters of this system are equal:
\begin{equation}
\alpha = -0.31084,   \mu_1 = 0.5,  N_1 = -1, N_3=-1,  A = 1.12,  B = -1.531.
\label{set2}
\end{equation}
We will call this set of parameters by the second set of parameters.

\begin{figure}
  \centerline{
    \includegraphics[width=0.55\textwidth]{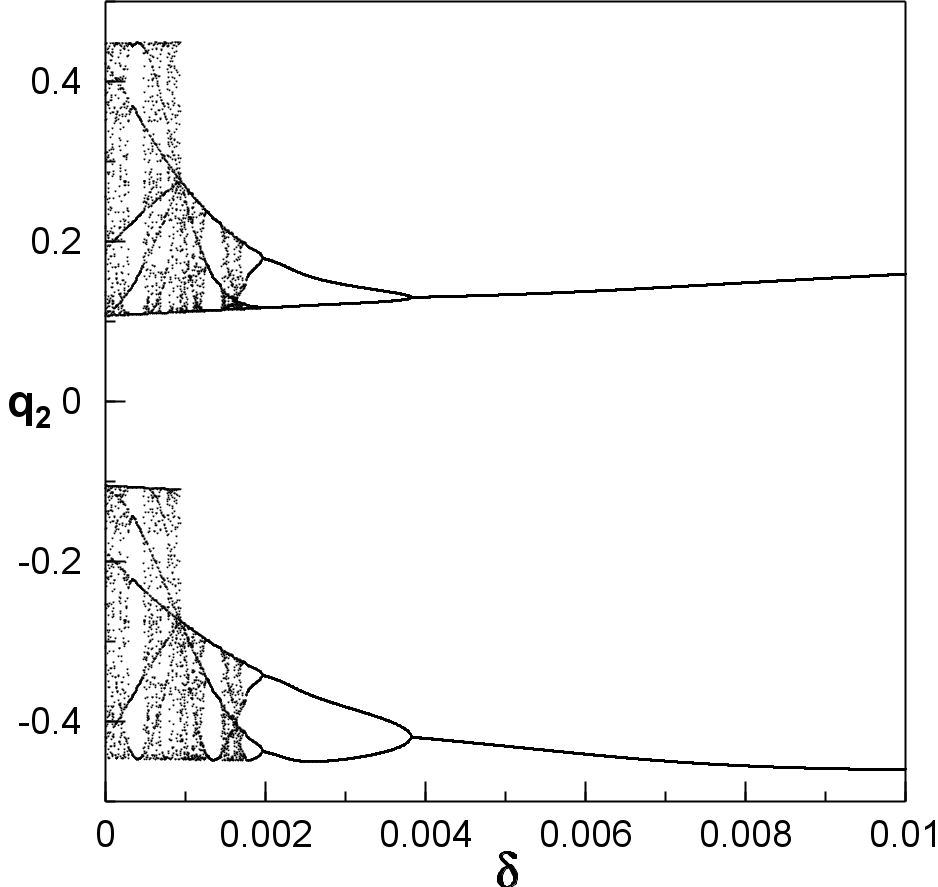}}
  \caption{Phase-parametric characteristic for second set of parameters }
  \label{fig:5}
\end{figure}

In Fig. \ref{fig:5} the phase-parametric characteristic of the system was constructed for the second set of parameters

In this phase-parametric characteristic, densely black areas correspond to chaotic attractors, areas consisting of individual branches of the bifurcation tree correspond to limit cycles.

\begin{figure}[ht]
\centering
\begin{minipage}{0.5\textwidth}
\centering
\includegraphics[width=\textwidth] {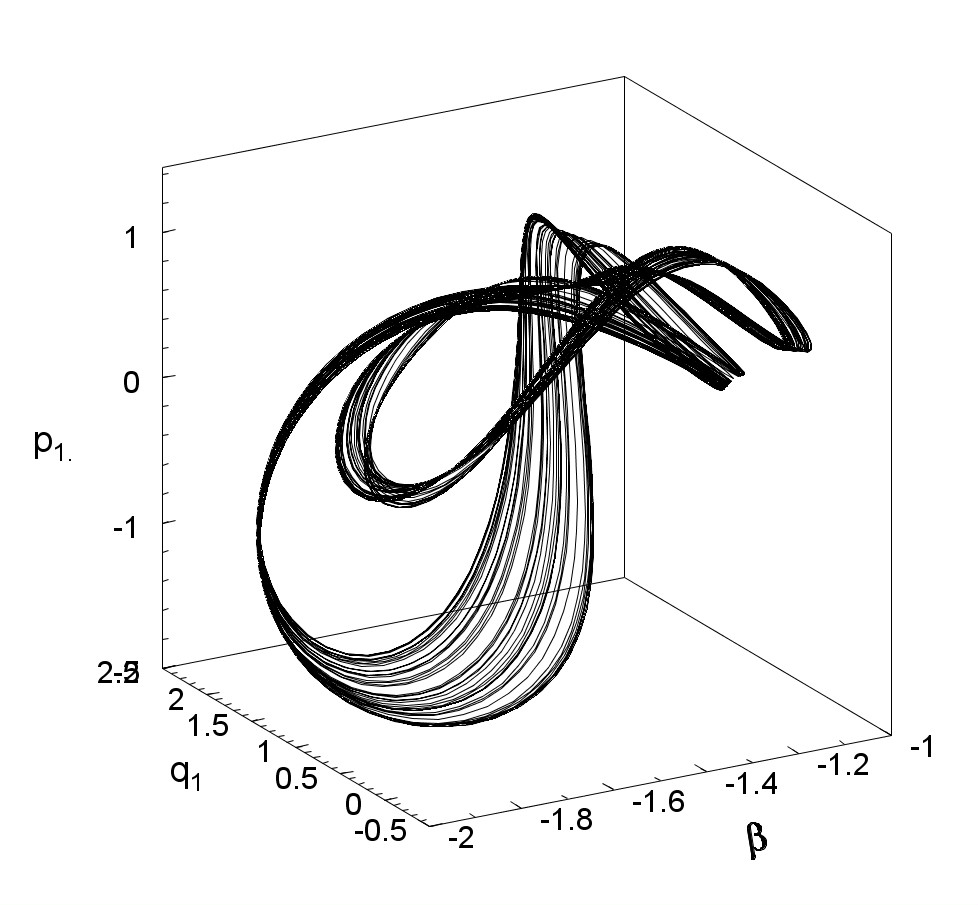}
(a)
\end{minipage}\hfill
\begin{minipage}{0.5\textwidth}
\centering
\includegraphics[width=\textwidth] {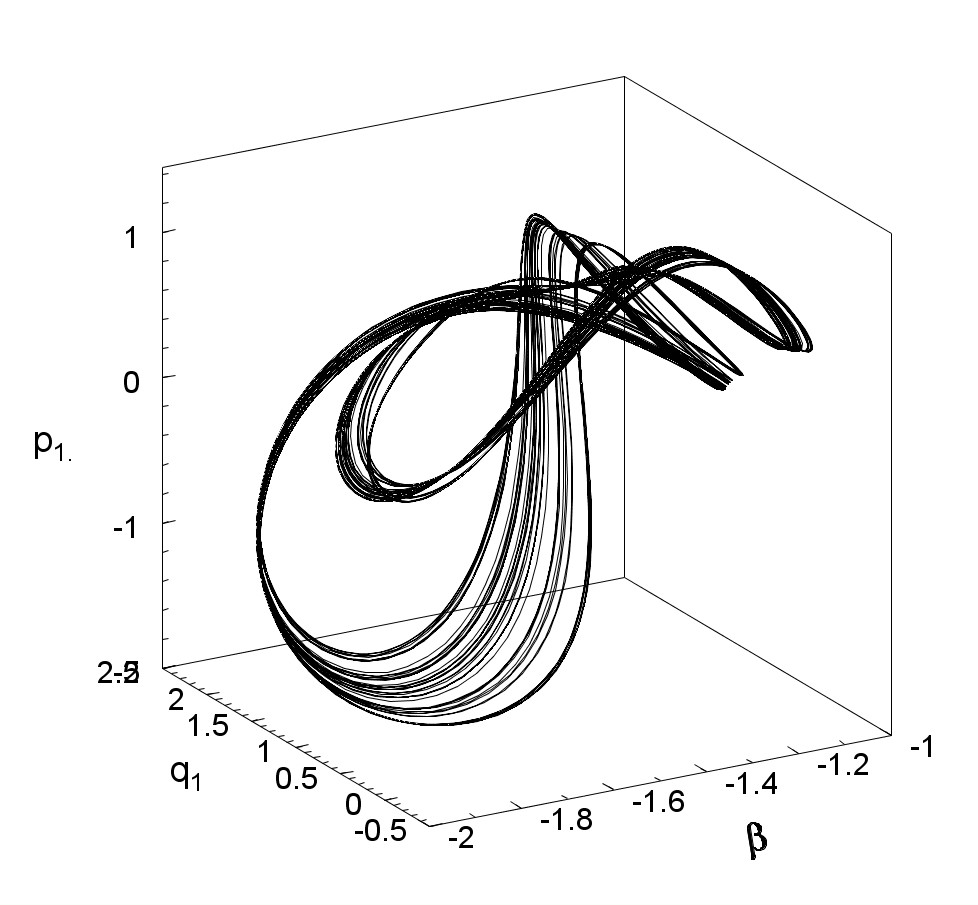}
(b)
\end{minipage}
\begin{minipage}{0.5\textwidth}
\centering
\includegraphics[width=\textwidth] {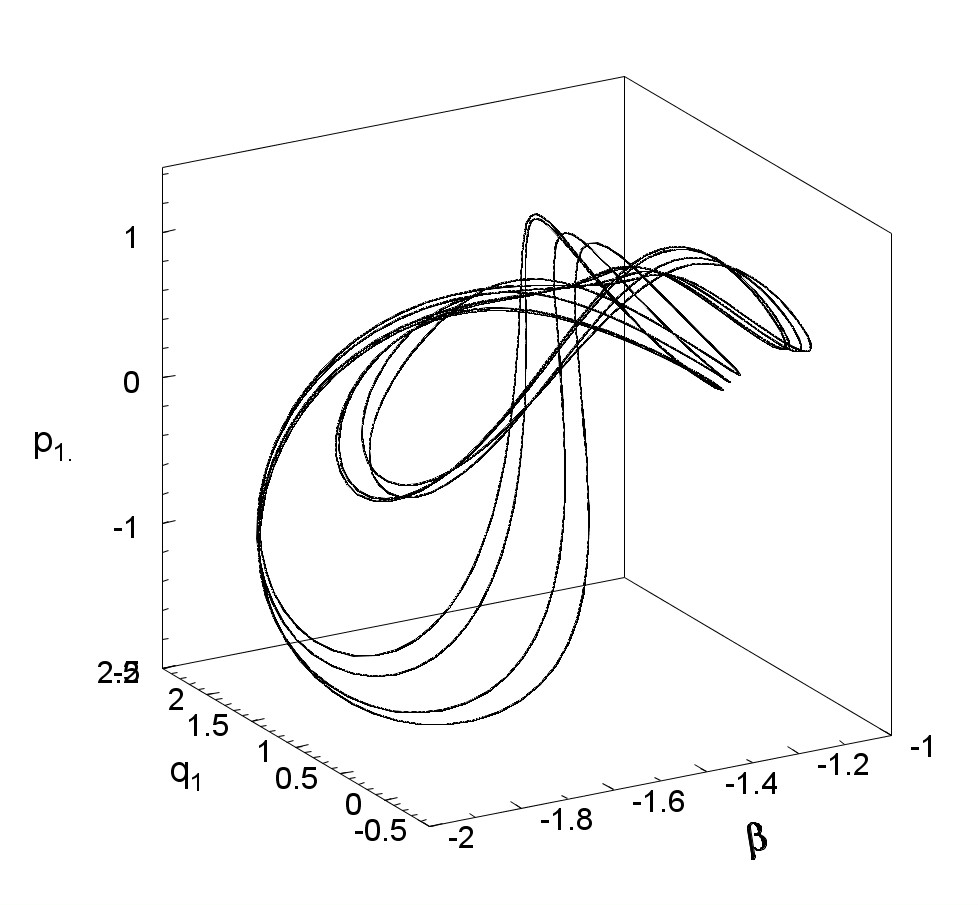}
(c)
\end{minipage}\hfill
\begin{minipage}{0.5\textwidth}
\centering
\includegraphics[width=\textwidth] {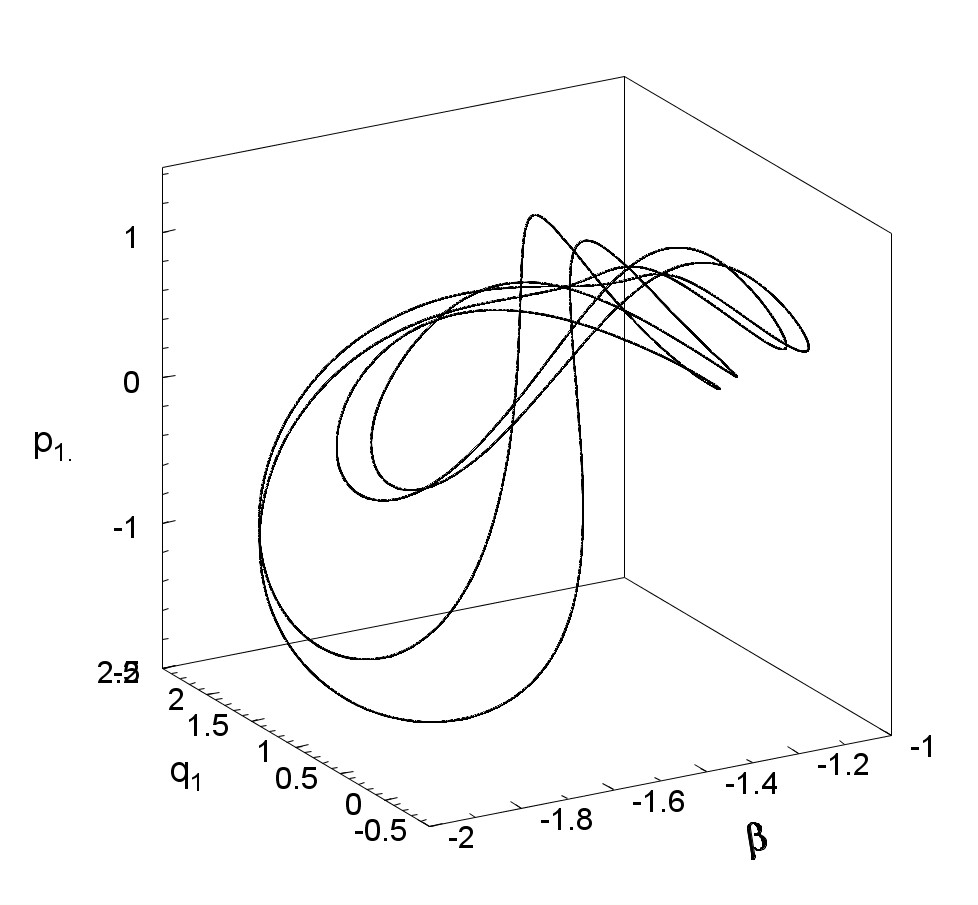}
(d)
\end{minipage}
\caption{Chaotic atttractors at $\delta=0$ (a), $\delta=0.001$ (b); limit cycle at $\delta=0.002$ (c), $\delta=0.004$ (d). }\label{fig:6}
\end{figure}

So in the absence of delay $(\delta=0) $ there is a chaotic attractor in the system (\ref{eq3}). The projection of the phase portrait of this attractor is plotted in Fig.\ref{fig:6} $(a)$. With a slight increase in the delay value $\delta=0.002$, the attractor of the system will still be a similar chaotic attractor (Fig. \ref{fig:6} (b)). However, with a further increase in delay, chaotic attractors disappear and limit cycles become attractors of the system. In Fig. \ref{fig:6} (c), (d) projections of phase portraits of two types of such cycles are shown.

\section{Conclusion}

Thus, the presence of a delay can significantly change the process of energy exchange between the excitation source (electric motor) and the oscillatory subsystem (tank partially filled with liquid). Even very small changes in delay values can generate chaos in the dynamic system under consideration. In other cases, the presence of delay can lead to the disappearance of chaotic attractors and the emergence of regular steady-state regimes.

\end{document}